\documentclass[amsmath,amssymb,twocolumn,aps,prd,floatfix]{revtex4-2}

\usepackage{graphicx}
\usepackage{dcolumn}
\usepackage{bm}
\usepackage{relsize}
\usepackage{MnSymbol}
\usepackage[utf8]{inputenc}
\usepackage{csquotes}

\begin{document}

\preprint{}

\title{\bf Transport coefficients of the heavy quark in the domain of the non-perturbative and non-eikonal gluon radiation}

\author{Surasree Mazumder}
\email{surasreem@iiserbpr.ac.in} 
\affiliation{Indian Institute of Science Education and Research, Berhampur}
\author{Natasha Sharma}%
\affiliation{Indian Institute of Science Education and Research, Berhampur}%
\author{Lokesh Kumar}
\affiliation{Panjab University}%

\date{\today}

\begin{abstract}

Drag and diffusion coefficients of the Heavy Quarks (HQs), such as charm and bottom, are one of the prime tools for discerning the properties of the deconfined QCD medium created in the heavy ion collisions experiments. The innate non-perturbative nature of the QCD medium renders it imperative to estimate the transport coefficients in that domain. The present work evaluates the drag and diffusion coefficients of the moving HQ interacting with the medium particles via two-body collisional and three-body radiative processes to the first order in opacity by employing Gribov mechanism for the non-perturbative regime. We proffer the latest results of the HQ transport coefficients computed for the non-perturbative and non-eikonal gluon radiation off the HQ. The calculations show significant increment of the transport coefficients with the increasing non-eikonality by juxtaposing the results with those of the perturbative and eikonal regions. We hope to shed fresh light towards explaining the experimental data on the nuclear modification factor, $R_{AA}$, the elliptic flow, $v_2$ of the HQ by advocating the importance of the non-eikonality of the gluon radiation off the HQ.

\end{abstract}

\maketitle

\section{\bf Introduction}
The formation of a strongly interacting colour deconfined matter comprising the light quarks/anti-quarks and gluons, is categorically established by colliding two heavy ions at relativistic speed in the experiments (Heavy Ion Collision (HIC) experiments)     conducted at the Relativistic Heavy Ion Collider (RHIC)~\cite{jadamsnpa2005,kadcoxnpa2005} and the Large Hadron Collider (LHC)~\cite{kaamodtprl2011,gaadprc2012,schatprc2014}. Plethora of studies have been performed throughout the global scientific community to delineate various characteristics of this QCD medium. One of the most efficient ways to characterise the microscopic and short-lived medium is to make use of different probes that are external to the bulk medium.

Heavy Quarks (HQs)~\cite{prinojpg2016,andronicepjc2016,aartsepja2017} such as charm and bottom that are produced in the primordial perturbative QCD hard collisions on a short time scale, $\tau_Q\sim 1/2M$~\cite{rappjhep2020}, $M$ being the mass of the heavy quark, emerge to be particularly adept probes to determine several salient features of the medium. Once the QCD medium is formed, it thermalises via QCD interactions among the constituent particles to become the Quark Gluon Plasma (QGP). Heavy Quarks, being produced even before the formation of the QGP, permeate the thermal medium by interacting with the partons and imprint the medium properties on itself until the freeze out time. Due to the heavier mass of the HQ than those of the constituent light partons and the temperature attained in the medium, thermal production of the HQ, moving as a Brownian particle in the medium, is highly improbable. Again, the number of HQ can be considered to be preserved throughout the evolution of the medium. It is possible to ascertain the bulk and transport properties of the QGP medium by observables like collective flow and energy loss connected to the HQ transport because of the relatively direct connections between the microscopic interactions of the HQ and the relevant observables.

The two processes that control the dynamics of the HQ in the medium constitute 1) $2\rightarrow 2$ collisions of the HQ with the medium particles and 2) $2\rightarrow 3$ gluon bremsstrahlung processes. While the first process principally measures the thermalisation of the HQ at lower momenta region, $p_Q\simeq \sqrt{3MT}$~\cite{rappjhep2020}, the second one becomes increasingly significant at higher HQ momenta. Medium-induced gluon radiation off the HQ has been comprehensively investigated in various literature~\cite{gossiauxjpg2010,armestoprd2004,zhangprl2004,djordjevicprc2008,dasprc2010,abirprc2016,duprd2018}. They described the process by soliciting the perturbative QCD (pQCD) technique concentrating on the domain of phase space where nonperturbative QCD (npQCD) effects remain subdued. 

Though pQCD can, successfully, explain the HQ experimental observables such as the nuclear modification factor ($R_{AA}$) and the elliptic flow ($v_2$) at higher HQ transverse momenta ($p_T$) regions, it fails to do so at the low to intermediate $p_T$ region of these data. Also, the physics near the critical temperature renders it necessary to understand the npQCD effects on the properties of the HQ dynamics. Although, lattice QCD describes the HQ dynamics most accurately, there are various limitations to use this approach for a full QCD medium, as yet. Therefore, studies~\cite{heesprc2006,grecoprl2008,aichelinprc2008,friesprc2012,alamjpg2013,songprc2015,beraudoepjc2015,scardinaprc2017,shaoepjc2018,xuprc2018,choprc2020} have been performed to estimate the non-perturbative transport coefficients of the HQ within the infra-structure of the pQCD. The T-matrix approach to calculate the HQ transport coefficients at low temperatures(closer to the critical temperature, $T_c$) has refined the method by making use of the interaction potential elicited from the finite-temperature lattice QCD studies and thus has taken a successful step towards computing the heavy quark transport coefficients in the low $p_T$ region~\cite{grecoprl2008,friesprc2012,liuprc2019}. Another approach has been proposed in Ref.~\cite{caoplb2023} in which a parameterized Cornell type HQ potential is assumed to work as an effective gluon propagator for the interaction of HQ with the medium partons to calculate HQ transport coefficients.
Some later studies estimated the diffusion coefficients~\cite{madniplb2023} of a non-relativistic HQ using and the energy loss of a fast moving parton~\cite{yusufprd2024} and  the Gribov-Zwanziger technique.

As the non-photonic single electron data suggest the importance of both the radiative as well as the collisional energy loss of HQ, it would be constructive to study the gluon radiation off the HQ in the npQCD regime. A premier study of the non-perturbative contributions of the radiative energy loss of the HQ has been performed in Ref.~\cite{rappjhep2020} following the T-matrix approach. A more recent literature~\cite{sumitprd2024} studies the non-perturbative transport coefficients of the HQ within the Gribov mechanism in the eikonal limit of gluon radiation off the heavy quark. In the present work, we have extended the study of the non-perturbative transport coefficients of the HQ beyond the eikonal approximation of the gluon radiation using the Gribov gluon propagator. The degree of the non-eikonality has been measured by the quantity, $\zeta=q_{\perp}/\sqrt{s}$, where $q_{\perp}$ is the magnitude of the transverse component of the exchanged momentum in the collision of the HQ with the thermal particles.  

This paper is organised as follows. In Sec.II, we discuss the approximations used, the notations adopted in the calculation, the non-perturbative Gribov gluon propagator and the calculation of the matrix elements of the interaction of the HQ with the thermal light quarks/anti-quarks and gluons. Then in the same section, we discuss the method to estimate the drag and the diffusion coefficients of the HQ moving with a velocity $\vec{v}$ inside QGP in the non-perturbative domain due to the non-eikonal radiation of gluons to the first order in opacity. Sec.III is dedicated to discuss and introspect the results and compare with the pQCD case as well as with the eikonal gluon radiation scenario. We summarise and conclude in Sec.IV.

\section{\bf Formalism}\label{formalism}
We discuss the methodology of the estimation of the medium induced energy loss due to gluon radiation from the HQ moving inside QGP in this section. Owing to the collision of the HQ with the medium particles, single or multiple gluons can be emitted from the process. We start by establishing the standard notations and the approximations adopted throughout the manuscript.
\subsection{\bf Approximations}\label{approx}
The rate of the energy loss by the HQ can be expanded in terms of the number of scattering events suffered by the HQ that is equivalent to an expansion in the powers of opacity. Opacity in a finite QGP medium can be defined as a product of the density of the medium with the transport cross section integrated along the path of the HQ. The lowest order in opacity amounts to one collisional interaction of the HQ with the medium, accompanied by the emission of a single gluon. Following are the approximations used in the present manuscript:
\begin{enumerate}
    \item Though we consider an infinite QGP medium in the present work, we stick to the same definition and expansion rule of opacity. We calculate the spectrum of the radiated gluon off the HQ up to the first order in opacity.

    \item The on-shell heavy quark is produced at the time, $t_0=-\infty$.
\end{enumerate}

The radiation spectrum of the emitted gluon from the collision of the HQ with the medium partons is, generally, obtained by scaling the Feynman amplitudes of $2\rightarrow 3$ radiative process ($|\mathcal{M}_{2\rightarrow 3}|^2$) by that of the $2\rightarrow 2$ collisional process ($|\mathcal{M}_{2\rightarrow 2}|^2$). $|\mathcal{M}_{2\rightarrow 2}|^2$ for the interaction of the HQ with the thermal light quarks/anti-quarks and gluons have been calculated and discussed vividly in plenty of studies~\cite{combridgenpb1979,svetitskyprd1988,braatenprd1991,mooreprc2005}. The analytical estimation of $|\mathcal{M}_{2\rightarrow 3}|^2$, however, involves various mathematical intricacies and kinematic approximations that are implemented in several respects. Most of the calculations of the gluon radiation spectrum in the lowest order in opacity have assumed soft, eikonal and collinear limits of the kinematic region. We briefly explain the terminology in order to establish the particular kinematic region we have considered in this work:
\begin{enumerate}
    \item Emission of soft gluons: the energy of the emitted gluon, $\omega\ll E$, energy of the HQ emitting the gluon.

    \item Emission of soft and collinear gluons: the transverse momentum of the radiated gluon, $k_{\perp} \ll \omega$.

    \item Eikonal propagation of the HQ: a) no recoil of the heavy quark takes place due to the collisional interaction, $E\gg q_{\perp}$, $q_{\perp}$ being the transverse momentum transfer of the collision process, b) No recoil of the HQ occurs due to the radiation of the gluon, $E\gg k_{\perp}$.
\end{enumerate}

Plenty of works have been performed in the scientific community that calculated the matrix elements of the radiative interaction of the HQ with the thermal partons via gluon emission in various kinematic limits including the one discussed above~\cite{mustafaplb1991,dokshitzerjpg1991,kharzeevplb2001,gyulassynpa2004,wicksnpa2007,abirprd2012,mazumderprd2014} and beyond~\cite{abirprd2013,abirprd2011,bhattacharyyaahep2016}. But, all of these studies discuss gluon radiation in the regime of the perturbative QCD. Fewer works~\cite{rappjhep2020,sumitprd2024} have discussed the radiation spectrum in the light of the non-perturbative QCD. We plan on estimating the diffusion coefficients of the moving HQ interacting through collision and radiation with the medium in the lowest order in opacity in an infinite QGP medium using the npQCD gluon propagator from the Gribov-Zwanziger formalism. We will relax the first eikonal approximation so that $E>q_{\perp}$ (instead of considering $E\gg q_{\perp}$). All other approximations will hold true.
\subsection{\bf Notations}
The Feynman amplitude, $|\mathcal{M}_{2\rightarrow 3}|^2$ will be calculated for the interaction of the HQ with the light quarks/anti-quarks described by $Q(k_1)+q/\bar{q}(k_2)\rightarrow Q(k_3)+q/\bar{q}(k_4)+g(k_5)$. In the approximation of the emission of the soft gluon, $\textit{i.e.}$ $\omega \ll E$, the radiative Feynman amplitude for the interaction of the HQ with the thermal gluons differ from that of the HQ with the light quarks only by a colour Casimir factor 9/4.
\begin{figure}[h]
\begin{center}
\includegraphics[width=0.48\textwidth]{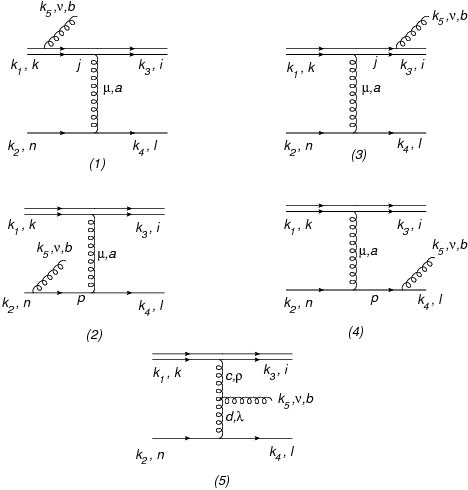}
\caption{t-channel diagram for single gluon emission. Double
line denotes heavy quarks. i, j, k, l, n, p are all quark colours. a, b, c, d are gluon colours and Greek letters denote Lorentz indices.}
\label{feyndia}
\end{center}
\end{figure}

Following the 4-momentum conservation of the 3-body interaction, $\textit{i.e.}$ $k_1+k_2=k_3+k_4+k_5$, let us define two sets of Mandelstam variables:
\begin{align}
    & s=(k_1+k_2)^2~~;~~s'=(k_3+k_4)^2\label{s}\\
    & t=(k_1-k_3)^2~~;~~t'=(k_2-k_4)^2\label{t}\\
    & u=(k_1-k_4)^2~~;~~u'=(k_2=k_3)^2
    \label{u}
\end{align}
with the constraint: $s+t+u+s'+t'+u'=4M^2$.

Dealing with the 3-body phase space is complicated. The soft gluon emission approximation enables one to convert the 3-body phase space into a 2-body one by taking the limit $k_5\rightarrow 0$~\cite{berendsplb1981}. This will lead to a further simplification of the Eqs.~\ref{s},\ref{t} and \ref{u} as follows:
\begin{equation}
    s\rightarrow s';~~t\rightarrow t';~~u\rightarrow u'\nonumber
\end{equation}
with the constraint, $s+t+u=2M^2$.

The Centre of Momentum (CoM) frame has been used in the present calculation with the following specification of the 4-momenta of the particles:
\begin{align}
    & k_1\equiv (E_1, \vec{0_{\perp}},k_{1z})~;~k_2\equiv (E_2, \vec{0_{\perp}},-k_{1z})\nonumber\\
    & k_3\equiv (E_3,\vec{q_{\perp}},k_{3z})~;~k_4\equiv (E_4, \vec{q_{\perp}},-k_{3z})\nonumber\\
    & k_5\equiv (\omega, \vec{k_{\perp}}=\omega \sin{\theta \hat{k_{\perp}}},\omega \cos{\theta})
\end{align}
where $\theta$ is the angle of radiation with the direction of propagation of the parent quark.

\subsection{\bf Gribov-Zwanziger gluon propagator}
One of the efficient ways to incorporate the effects of the npQCD within the framework of the standard pQCD calculations is the utilization of the prescription given by Gribov-Zwanziger~\cite{gribovnpb1978,zwanzigernpb1989,zwanzigerprl2005,zwanzigerprd2007} for the gluon propagator. The general-gauge gluon propagator is written as
\begin{equation}
    D^{\mu\nu}(Q)=\left[\delta^{\mu\nu}-(1-\xi)\frac{Q^{\mu}Q^{\nu}}{Q^2}\right]\frac{Q^2}{Q^4+\gamma_G^4}.
    \label{gribov}
\end{equation}
Here, $Q^{\mu}$ is the four-momentum of the gluon, $\xi$ is the gauge parameter and $\gamma_G$ is the Gribov parameter. One can assert the justification of using Gribov prescription in the deconfined matter from various literatures~\cite{zwanzigerprl2005,zwanzigerprd2007}. It can be argued from the theoretical vantage point~\cite{fukushimaprd2013} and the numerical results from lattice QCD~\cite{borsanyijhep2012} that the Gribov gluon propagator can be used in the domain of this work, safely.

The Gribov parameter, $\gamma_G$ can be obtained by solving a self-consistent gap equation~\cite{gribovnpb1978,zwanzigernpb1989} defined for infinite order of loop. Nevertheless, at asymptotically high temperatures and at the order of one loop, $\gamma_G\sim g^2T$ which behaves like magnetic scale. Even in the theory of pQCD, one can, in principle, include $\gamma_G$ on top of the Debye mass which only gives the electric scale ($~gT$) in order to better regulate the infrared divergence. As we are mainly interested to observe the effect of the npQCD on the gluon radiation spectrum, we take $\gamma_G(T)$ to be a general function of temperature fitted from the lattice equation of state as depicted by the authors of Ref.~\cite{jaiswalplb2020}.

\subsection{\bf Calculation of the radiative Feynman amplitudes, $|\mathcal{M}_{2\rightarrow 3}|^2$}
The Feynman amplitudes for the process $Q(k_1)+q/\bar{q}(k_2)\rightarrow Q(k_3)+q/\bar{q}(k_4)$ as depicted in Fig.\ref{feyndia} are
\begin{align}
    & \mathcal{M}_{11}^2=\frac{128}{27}g^6\frac{s^2}{t^2}\frac{1}{k_{\perp}^2}\left(-\frac{1}{\tan^2{\theta/2}}\right)\mathcal{T}^2\nonumber\\
    &\times\left[T_M^2+\frac{\mathcal{C}_{11}}{\left(1-T_M^2\right)^2}\right]\mathcal{N}_p\label{m11}\\
    & \mathcal{M}_{13}^2=\frac{128}{27}g^6\frac{s^2}{t^2}\frac{1}{k_{\perp}^2}\frac{1}{4\mathcal{F}_{35}}\left(-\frac{1}{\tan^2{\theta/2}}\right)\mathcal{T}^2\nonumber\\
    & \times\left[T_M^2-\frac{\mathcal{C}_{13}}{\left(1-T_M^2\right)^2}\right]\mathcal{N}_p\label{m13}\\
    & \mathcal{M}_{12}^2=\frac{128}{27}g^6\frac{s^2}{t^2}\frac{1}{k_{\perp}^2}\frac{1}{4}\left(1-T_M^2\right)\mathcal{T}\nonumber\\
    & \times\left[1-\frac{\mathcal{C}_{12}}{\left(1-T_M^2\right)^3}\right]\mathcal{N}_p\label{m12}\\
    & \mathcal{M}_{23}^2=\frac{128}{27}g^6\frac{s^2}{t^2}\frac{1}{k_{\perp}^2}\frac{7}{8\mathcal{F}_{35}}\left(1-T_M^2\right)\mathcal{T}\nonumber\\
    &\times\left[1+\frac{\mathcal{C}_{23}}{\left(1-T_M^2\right)^3}\right]\mathcal{N}_p\label{m23}\\
    & \mathcal{M}_{24}^2=\frac{128}{27}g^6\frac{s^2}{t^2}\frac{1}{k_{\perp}^2}\frac{1}{8}\frac{t}{s}\frac{\tan^2{\theta/2}}{\mathcal{F}_{45}}\mathcal{N}_p\label{m24}\\
    & \mathcal{M}_{33}^2=\frac{\mathcal{M}_{11}^2}{\mathcal{F}_{35}^2}\label{m33}\\
    & \mathcal{M}_{14}^2=\frac{\mathcal{M}_{23}^2\mathcal{F}_{35}}{\mathcal{F}_{45}}\label{m14}\\
    & \mathcal{M}_{34}^2=\frac{\mathcal{M}_{12}^2}{\mathcal{F}_{35}\mathcal{F}_{45}}\label{m34}\\
    & \mathcal{M}_{22}^2=\mathcal{M}_{44}^2=\mathcal{M}_{55}^2=0\label{m224455}\\
    & \mathcal{M}_{15}^2=\mathcal{M}_{25}^2=\mathcal{M}_{35}^2=\mathcal{M}_{45}^2=0\label{m15253545},
\end{align}
where the notations denoted by $T_M$, $\mathcal{T}$, $\mathcal{C}_{11}$, $\mathcal{C}_{13}$, $\mathcal{C}_{12}$, $\mathcal{C}_{23}$, $\mathcal{C}_{24}$, $\mathcal{F}_{35}$ and $\mathcal{F}_{45}$ are given, retaining terms which contain $\frac{t}{s}$ (because of the approximation $E_1>q_{\perp}$), as:
\begin{align}
    & T_M=\frac{M}{\sqrt{s}}\\
    & \mathcal{T}=\frac{1-T_M^2}{1+\frac{T_M^2}{\tan^2{\theta/2}}}\label{T}\\
    & \mathcal{C}_{11}=T_M^2\frac{t}{s}\left(1+\frac{t}{2s}\right)\label{c11}\\
    & \mathcal{C}_{13}=2\frac{t}{s}-T_M^2\frac{t}{2s}\label{c13}\\
    & \mathcal{C}_{12}=T_M^2\frac{t}{s}-\frac{t}{s}\label{c12}\\
    & \mathcal{C}_{23}=\frac{2t}{s}-3T_M^2\frac{t}{s}\label{c23}\\
    & \mathcal{F}_{35}=1+\frac{T}{\tan{\theta/2}\left(1+\frac{T_M^2}{\tan^2{\theta/2}}\right)}\label{f35}\\
    & T=\cot{\theta}\left[\left(1-T_M^2\right)-\sqrt{\left(1-T_M^2\right)^2-4\left(\frac{q_{\perp}}{\sqrt{s}}\right)^2}\right]\nonumber\\
    & -2\left(\frac{q_{\perp}}{\sqrt{s}}\right)\\
    & \mathcal{F}_{45}=1-\frac{T}{\left(1-T_M^2\right)\cot{\theta/2}}\label{f45}.
    \end{align}
The non-perturbative effect is portrayed by the term $\mathcal{N}_p$, given by
\begin{equation}
    \mathcal{N}_p=\frac{t^4}{(t^2+\gamma_G^4)^2}
\end{equation}
The Feynman amplitudes expressed in Eqs.~[\ref{m11}-\ref{m15253545}] have been calculated using the Gribov gluon propagator (Eq.~[\ref{gribov}]) taking the Feynman gauge that is $\xi=1$. In the CoM frame,
\begin{align}
    & \frac{t}{s}=-\frac{q_{\perp}^2}{s}-\frac{1}{4}\bigg[\left(1-T_M^2\right)\nonumber\\
   & -\sqrt{\left(1-T_M^2\right)^2-4\left(\frac{q_{\perp}}{\sqrt{s}}\right)^2}\bigg]^2\\
   & s=2E_1^2-M^2+2E_1^2\sqrt{E_1^1-M^2}
\end{align}
Total gauge-invariant squared matrix elements or the Feynman amplitude for the interaction of the HQ with the light quarks/anti-quarks is the summation of all the Feynman amplitudes of the five diagrams depicted in Fig.~\ref{feyndia} and their interference terms:
\begin{equation}
    |\mathcal{M}_{Qq\rightarrow Qqg}|^2=\sum_{i\ge j}\mathcal{M}_{ij}^2,
\end{equation}
in which $\textit{i}$ and $\textit{j}$ run from 1 to 5 and $\mathcal{M}_{ij}=\mathcal{M}_i\mathcal{M}_j^{*}$, $\mathcal{M}_i$ being the matrix element of the \textit{i$^{th}$} diagram of the Fig.~\ref{feyndia}. The total squared matrix element can be concisely written as the following:
\begin{equation}
    |\mathcal{M}_{Qq\rightarrow Qqg}|^2=\frac{16}{3}g^2\frac{1}{\omega^2\sin^2{\theta}}|\mathcal{M}_{Qq\rightarrow Qq}|^2\mathcal{W}(x,k_{\perp}^2)\label{mQq},
\end{equation}
where, the gluon rapidity, $\eta=-\ln(\tan{\theta/2})$, light cone variable, $x=k_{\perp}e^{\eta}/\sqrt{s}$, the effective gluon radiation spectrum
\begin{align}
    & \mathcal{W}(x,k_{\perp}^2)=\mathlarger{\mathlarger{\sum}}_{n=2,1,0}\mathcal{C}_ne^{2(n-1)\eta}\left(\frac{k_{\perp}^2}{k_{\perp}^2+x^2M^2}\right)^n.
    \label{radspec}
\end{align}
The coefficients $\mathcal{C}_n~'s$ are given by
\begin{align}
    & \mathcal{C}_0=\frac{1}{8\mathcal{F}_{45}\left(1-T_M^2\right)^2}\frac{t}{s},\label{c0}\\
    & \mathcal{C}_1=\frac{C}{4}\left(1+\frac{1}{\mathcal{F}_{35}\mathcal{F}_{45}}\right)+\frac{7}{8}D\left(\frac{1}{\mathcal{F}_{35}}+\frac{1}{\mathcal{F}_{45}}\right),\label{c1}\\
    & \mathcal{C}_2=-\left(A+\frac{A}{\mathcal{F}_{35}^2}+\frac{B}{4\mathcal{F}_{35}}\right),\label{c2}
\end{align}
with
\begin{align}
    & A=T_M^2+\frac{\mathcal{C}_{11}}{\left(1-T_M^2\right)^2},\label{a}\\
    & B=T_M^2-\frac{\mathcal{C}_{13}}{\left(1-T_M^2\right)^2},\label{b}\\
    & C=1-\frac{\mathcal{C}_{12}}{\left(1-T_M^2\right)^3},\label{c}\\
    & D=1+\frac{\mathcal{C}_{23}}{\left(1-T_M^2\right)^3}\label{d}.
\end{align}

Squared matrix element for the process $Qg\rightarrow Qgg$ is
\begin{equation}
  |\mathcal{M}_{qg\rightarrow Qgg}|^2=\frac{9}{4}|\mathcal{M}_{Qq\rightarrow Qqg}|^2.
  \label{mQg}
\end{equation}

One might refer to Ref.\cite{bhattacharyyaahep2016} in which the authors explored the spectrum of the non-eikonal gluon radiation off the HQ, for more details. It can easily be corroborated that in the limit, $\zeta=0$, \textit{i.e.} in the eikonal limit, all these expressions reduce to the eikonal results of the Ref.~\cite{abirprd2012}.

\subsection{\bf Non-perturbative drag and diffusion coefficients of the HQ for non-eikonal gluon radiation}
As mentioned earlier, the motion of the heavy quark can be described by the equations of the Brownian motion. One of the alternative approaches to describe such a motion is to write down the relevant Fokker-Planck equation for the HQ moving with an initial momentum, $\vec{p}$ and define the subsequent transport coefficients~\cite{svetitskyprd1988}. The form of the Fokker Planck Equation, which is the approximated rendition of the general Boltzmann transport equation in the limit of the soft momentum exchange between the HQ and the medium particles, is
\begin{equation}
    \frac{\partial f_Q}{\partial t}=\frac{\partial}{\partial k_{1i}}\left[A_i(\vec{k_1})f_Q+\frac{\partial}{\partial k_{1j}}[B_{ij}(\vec{k_1})f_Q]\right].
    \label{FPE}
\end{equation}
In Eq.~[\ref{FPE}], $f_Q$ is the distribution function of the HQ, $A_i$ is related to the drag of the HQ caused by the interaction with the medium and $B_{ij}$ is the HQ diffusion tensor generated due to the random kicks from the medium. For an isotropic medium,
\begin{align}
    & A_i(\vec{k_1})=\eta(\vec{k_1})k_{1i}\label{drag}\\
    & B_{ij}(\vec{k_1})=\frac{k_{1i}k_{1j}}{|\vec{k_1}|^2}\kappa_L(\vec{k_1})+\left(\delta_{ij}-\frac{k_{1i}k_{1j}}{|\vec{k_1}|^2}\right)\kappa_T(\vec{k_1}),
    \label{diffusion}
\end{align}
where, $\eta(\vec{k_1})$, $\kappa_L(\vec{k_1})$ and $\kappa_T(\vec{k_1})$ are the drag, longitudinal diffusion and transverse diffusion coefficients of the HQ, respectively. In the following two subsequent subsections we discuss the methodology of the estimation of these transport coefficients of the HQ for collisional and radiative interactions, respectively adhering to the quantum field theoretical approach.

\subsubsection{\bf Collisional transport coefficients of the HQ}
For the two-body process, $Q(k_1)+q/\bar{q}/g(k_2)\rightarrow Q(k_3)+q/\bar{q}/g(k_4)$, the drag and diffusion coefficients of the HQ~\cite{svetitskyprd1988} can be evaluated by solving the following:
\begin{align}
    &A_i(\vec{k_1})=\frac{1}{2E_1}\left(\mathlarger{\mathlarger{\prod}}_{i=2,3,4}\int\frac{d^3\vec{k_i}}{(2\pi)^32E_i}\right)\delta^4(k_1+k_2-k_3-k_4)\nonumber\\
    & (2\pi)^4(k_1-k_3)_i\mathcal{M}_{col}\nonumber\\
    & =\llangle(k_1-k_3)_i\rrangle,
    \label{Ai}
\end{align}
and
\begin{equation}
    B_{ij}(\vec{k_1})=\frac{1}{2}\llangle(k_1-k_3)_i(k_1-k_3)_j\rrangle,
    \label{Bij}
\end{equation}
where, 
\begin{eqnarray}
\mathcal{M}_{col}&=&\bigg[|\mathcal{M}_{Qq\rightarrow Qq}|^2f(k_2)[1-f(k_4)]\nonumber\\
&+&|\mathcal{M}_{Qg\rightarrow Qg}|^2f(k_2)[1+f(k_4)]\bigg]
\end{eqnarray}
By virtue of the Eqs.~[\ref{drag}-\ref{Bij}], one can write:
\begin{align}
    & \eta=\llangle1\rrangle-\frac{\llangle\vec{k_1}\cdot\vec{k_3}\rrangle}{|\vec{k_1}|^2}\label{eta}\\
    & \kappa_L=\frac{1}{2}\left[\frac{\llangle(\vec{k_1}\cdot\vec{k_3})^2\rrangle}{|\vec{k_1}|^2}-2\llangle\vec{k_1}\cdot\vec{k_3}\rrangle+|\vec{k_1}|^2\llangle1\rrangle\right]\label{kappaL}\\
    & \kappa_T=\frac{1}{4}\left[\llangle|\vec{k_3}|^2\rrangle-\frac{\llangle(\vec{k_1}\cdot\vec{k_3})^2\rrangle}{|\vec{k_1}|^2}\right].\label{kappaT}
\end{align}
The phase space integration,
\begin{align}
    & I=\frac{1}{2E_1}\mathlarger{\mathlarger{\prod}}_{i=2,3,4}\int\frac{d^3\vec{k_i}}{(2\pi)^32E_i}(2\pi)^4\delta^4(k_1+k_2-k_3-k_4)\label{I}\nonumber\\
\end{align}
can be evaluated by aligning the initial HQ momentum, $\vec{k_1}$ along the z-axis. The $|\vec{k_3}|$ integration can be performed using the 3-dimensional delta function. In the soft momentum transfer limit, $\textit{i.e.}$ $E_3\equiv E_1-\vec{v}\cdot\vec{q}$, $\vec{q}=\vec{k_1}-\vec{k_3}=\vec{k_4}-\vec{k_2}$ being the 3-momentum transfer between the HQ and the medium particle, the energy delta function becomes $\delta(\omega-\vec{v}\cdot\vec{q})$. Let us denote the angles made by $\vec{k_2}$ and $\vec{k_4}$ with the z-axis by $\theta_2$ and $\theta_4$, respectively. If the corresponding azimuthal angles are designated as $\phi_2$ and $\phi_4$, one can write down Eq.~[\ref{I}] as:
\begin{align}
    & I=\frac{1}{(2\pi)^516E_1^2}\int_{0}^{\infty}\omega d\omega\int_{\frac{\omega}{1+v}}^{\frac{\omega}{1-v}}d|\vec{k_2}|\int d|\vec{k_4}|\nonumber\\
    & \int_{0}^{2\pi}d \phi_2\int_{0}^{2\pi}d \phi_4.
\end{align}
The Ref.~\cite{mooreprc2005} discusses the detailed analysis of the phase space integration.

\subsubsection{\bf Radiative transport coefficients of the HQ}
For the single gluon emission process, $Q(k_1)+q/\bar{q}/g(k_2)\rightarrow Q(k_3)+q/\bar{q}/g(k_4)+g(k_5)$,
the generic transport coefficient, $X(\vec{k_1})$ can be expressed as~\cite{mazumderprd2014}:
\begin{align}
    & \llangle X(\vec{k_1})\rrangle_{rad}=\frac{1}{2E_1}\left(\mathlarger{\mathlarger{\prod}}_{i=2,3,4,5}\int\frac{d^3\vec{k_i}}{(2\pi)^32E_i}\right)\nonumber\\
    & \times(2\pi)^4\delta^4(k_1+k_2-k_3-k_4)[1+f(k_5)]\nonumber\\
    & \times X(\vec{k_1})\theta_1(\tau_s-\tau_F)\theta_2(E_1-E_5)\mathcal{M}_{rad},
    \label{Xrad}
\end{align}
where,
\begin{eqnarray}
\mathcal{M}_{rad}&=&\bigg[|\mathcal{M}_{Qq\rightarrow Qqg}|^2f(k_2)[1-f(k_4)]\nonumber\\
&+&|\mathcal{M}_{Qg\rightarrow Qgg}|^2f(k_2)[1+f(k_4)]\bigg].
\end{eqnarray}
$\theta_1$ and $\theta_2$ in Eq.~[\ref{Xrad}] represent two different kinematic conditions stemmed from two different assumptions:i) the scattering time of the HQ with the medium, $\tau_s$ must be greater than the formation time of the radiated gluon, $\tau_F$, which is called the Landau-Pomeronchuk-Migdal (LPM) effect~\cite{wangprd1995,kleinrpm1999} and ii) the energy of the radiated gluon cannot be greater than the energy of the incident HQ. From the condition of the LPM effect, we extract the minimum bound on $k_{\perp}$ because $\theta_1(\tau_s-\tau_F)$ amounts to
\begin{equation}
    \tau_s=\Gamma^{-1}>\tau_F=\frac{\cosh{\eta}}{k_{\perp}},
\end{equation}
where, $\Gamma$ is the interaction rate of the HQ with the medium and which sets $k_{\perp}^{min}=\Gamma \cosh{\eta}$. From the second theta-function, one can get the maximum bound on $k_{\perp}$ as the following:
\begin{equation}
    E_1>E_5=k_{\perp}\cosh{\eta},~~k_{\perp}^{max}=\frac{E_1}{\cosh{\eta}}.
\end{equation}
In the kinematic region as described in Sec.II(B), $\textit{i.e.}$, $k_5\rightarrow 0$, transport coefficient for the radiative interaction can be expressed in terms of that of the collisional interaction as follows:
\begin{align}
    & \llangle X(\vec{k_1})\rrangle_{rad}=\llangle X(\vec{k_1})\rrangle_{col}\times\frac{16}{3}g^2\int\frac{d^3|\vec{k_5}|}{(2\pi)^32E_5}\frac{1}{k_{\perp}^2}\nonumber\\
    & \times \mathcal{W}(x,k_{\perp})[1+f(k_5)]\theta_1(\tau_s-\tau_F)\theta_2(E_1-E_5),
    \label{Xrad2}
\end{align}
where, $\mathcal{W}(x,k_{\perp})$ is taken from Eq.~[\ref{radspec}].
Here, $d^3|\vec{k_5}|=d^2k_{\perp}dk_z$, $E_5=k_{\perp}\cosh{\eta}$ and $k_z=k_{\perp}\sinh{\eta}$.

\section{\bf Results and Discussion}
\begin{figure}[h!]
 \centering
  \includegraphics[width=0.48\textwidth,angle=0]{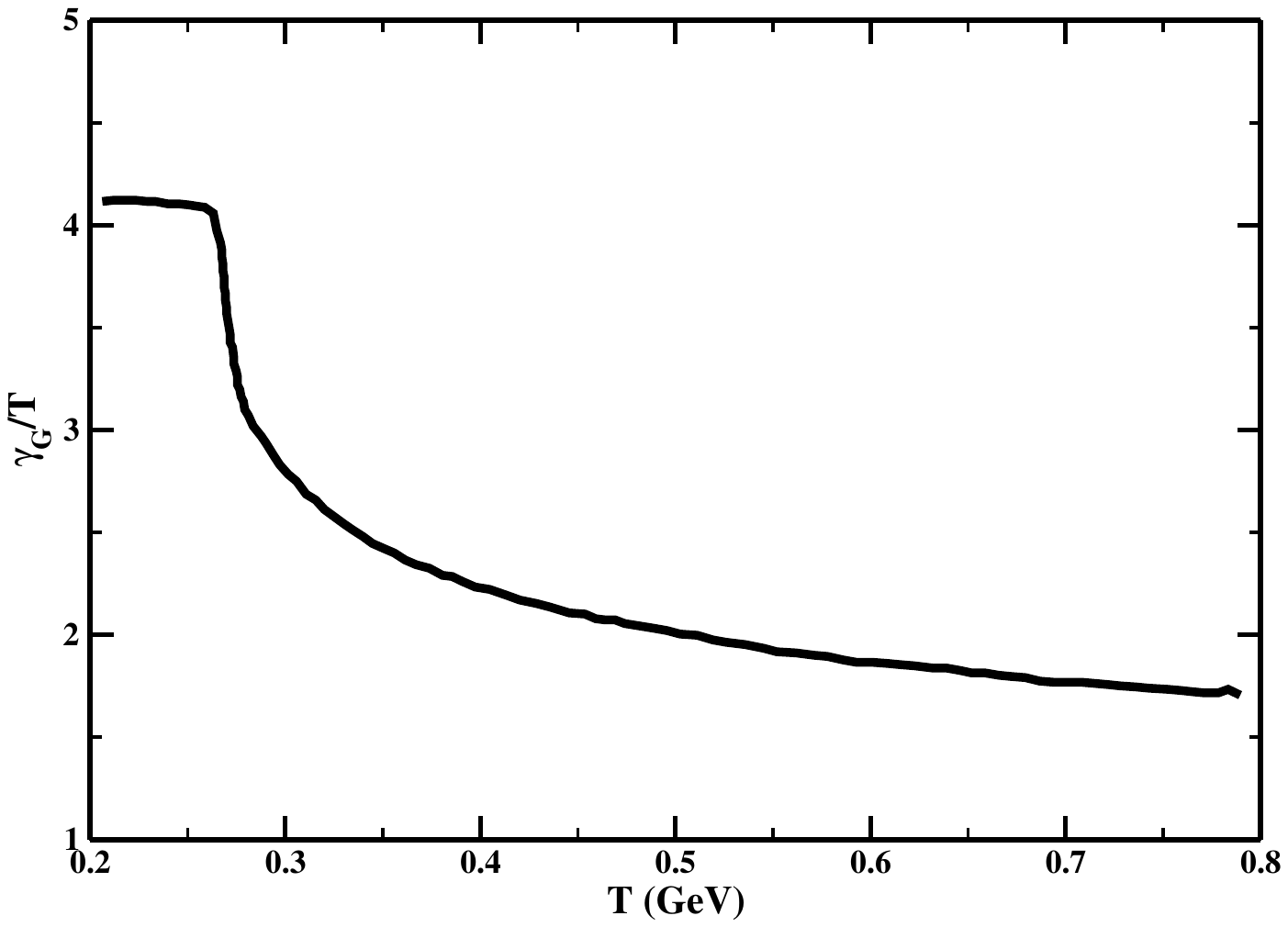}
   \caption{Variation of the scaled Gribov parameter, $\gamma_G/T$, with T  as discussed in the Refs.~\cite{sumitprd2024,jaiswalplb2020}.}
    \label{fig:f0}
\end{figure}

The estimation of the transport coefficients of the charm quark in the non-perturbative regime entails an appropriate parametrisation of the Gribov parameter, $\gamma_G$ as a function of temperature of the medium. Authors of Ref.~\cite{jaiswalplb2020} have obtained the requisite temperature dependence of the scaled Gribov parameter, \textit{i.e.} $\gamma_G/T$ by matching the thermodynamics of the medium with the pure lattice data~\cite{borsanyijhep2012}. As manifested in the Fig.~\ref{fig:f0}, the temperature dependence of $\gamma_G$ has been used in all the numerical determination of the various transport coefficients of the charm quark. The principal objective of the present work is to indicate the consequences of the non-eikonality discussed in Sec.\ref{approx} on the various transport coefficients of the charm quark, evaluated in the non-perturbative regime.

\begin{figure}[h!]
 \centering
  \includegraphics[width=0.48\textwidth,angle=0]{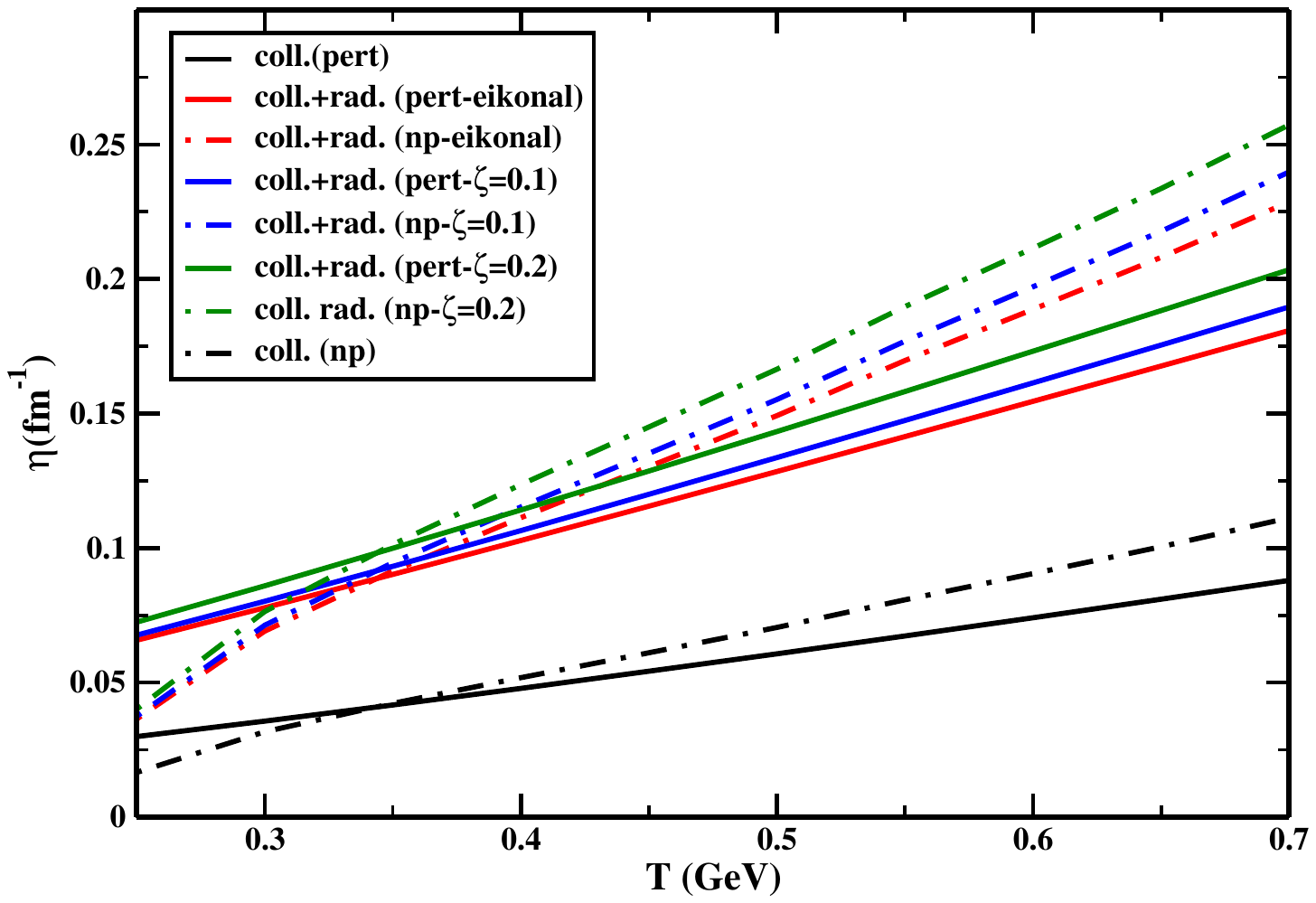}
   \caption{Variation of the drag coefficient, $\eta$, of the charm quark having momentum 5 GeV with temperature of the medium for various eikonalities. Here, the terms \enquote{pert.}, \enquote{np}, \enquote{coll} and \enquote{rad} indicate perturbative, non-perturbative, collisional and radiative respectively. }
    \label{fig:f1}
\end{figure}

Figure~\ref{fig:f1} portrays the temperature dependence of the drag coefficient, $\eta$, of the charm quark having mass 1.3 GeV and momentum 5 GeV for various degrees of eikonality, $\zeta$, of the radiative process. It can be observed that both the collisional and the radiative processes share similar contributions to the drag coefficient of charm. The effect of the non-perturbative physics tends to prevail over the perturbative one due to the temperature dependence of the Gribov parameter, $\gamma_G$. The dominance of the radiative process is seen to have increased with the increasing temperature for the perturbative as well as the non-perturbative cases. The inclusion of the non-perturbative effect through the Gribov mechanism increases the magnitude of the drag coefficient of the charm quark. The Gribov parameter encodes the information that the medium is strongly coupled so that the interactions of the charm quark will be stronger leading to a rise in the magnitude of the drag coefficient.  

\begin{figure}[ht!]
 \centering
  \includegraphics[width=0.48\textwidth,angle=0]{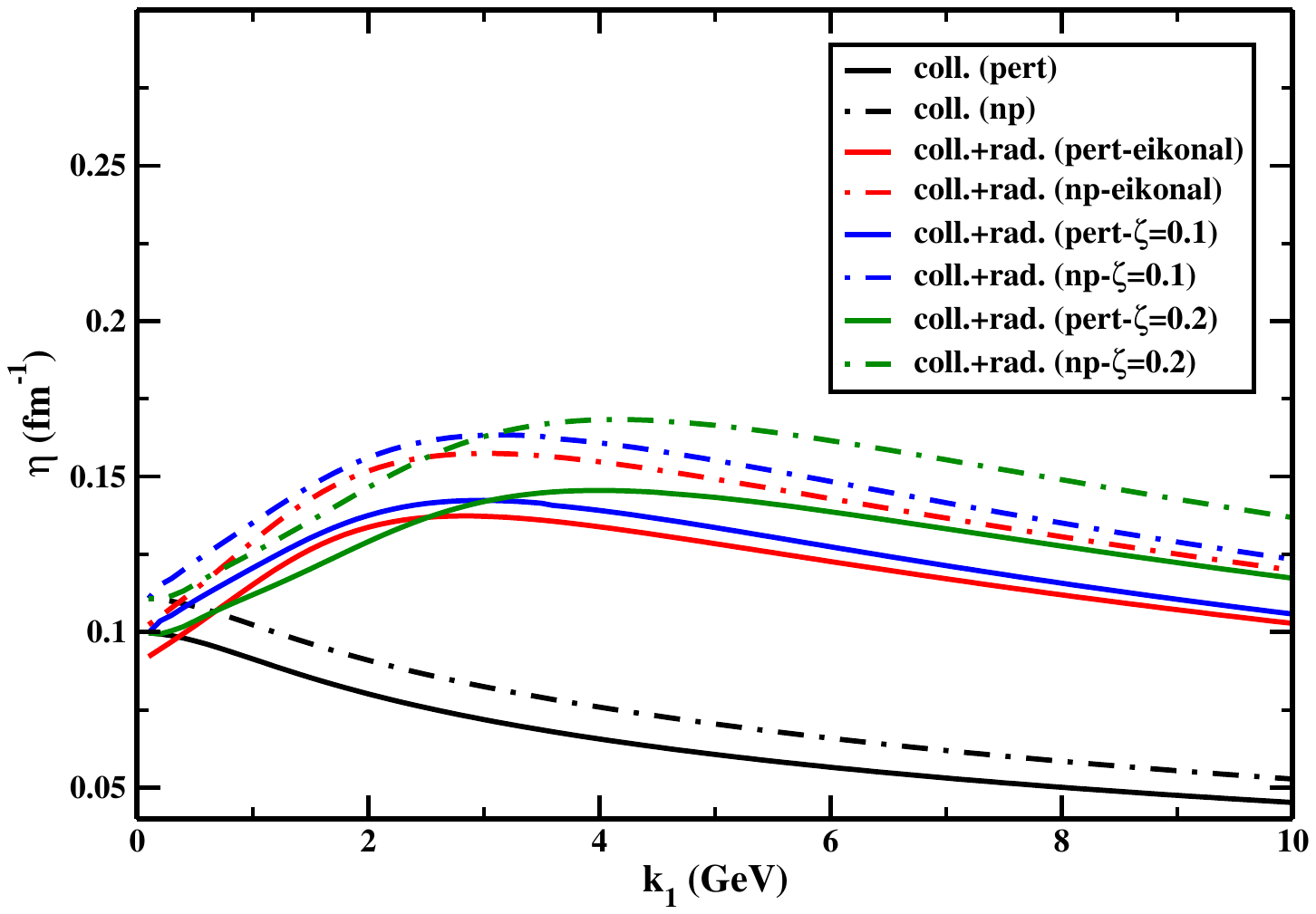}
   \caption{Variation of the drag coefficient, $\eta$, of the charm quark with momentum at a temperature 500 MeV for different eikonalities, $\zeta$.}
    \label{fig:f2}
\end{figure}

The variation of the drag coefficient, $\eta$, of charm with the charm momentum at a temperature of 500 MeV is depicted in Fig.~\ref{fig:f2}. Similar to the previous figure, we see that the drag is increasing as we encompass the effects of the non-perturbative QCD. One can also stipulate that the increment in the degree of non-eikonality enhances the drag coefficient of the charm quark. As the non-eikonality, $\zeta$, increases, the probability for the charm quark to deflect from the straight path after each collision increases. The more the charm quark deflects, the probability to find more medium particles with which it can interact increases, too, leading to an enhanced drag coefficient. 

\begin{figure}[ht!]
 \centering
  \includegraphics[width=0.48\textwidth,angle=0]{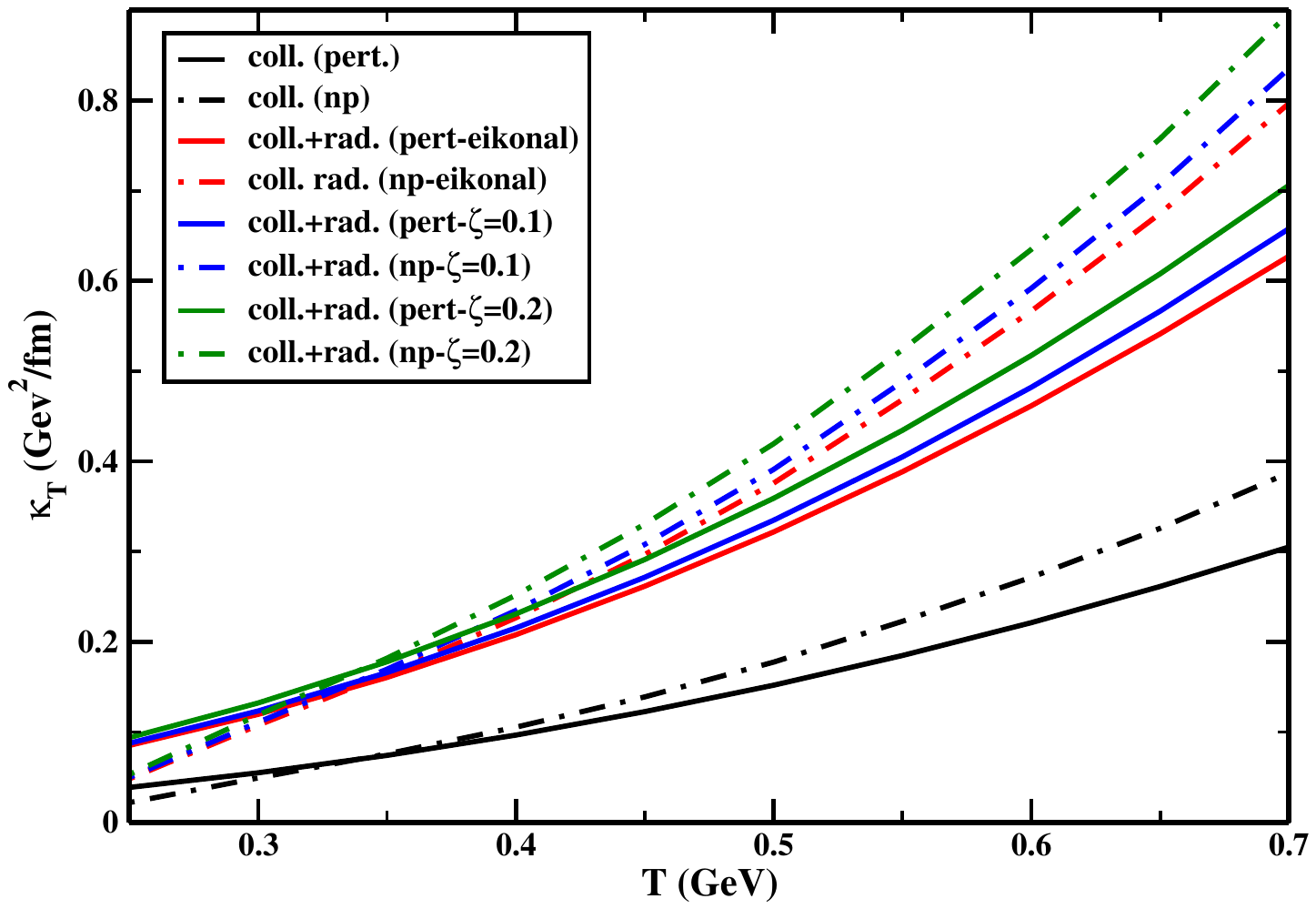}
   \caption{Variation of the transverse diffusion coefficient, $\kappa_T$, of charm having momentum 5 GeV with temperature of the medium for different eikonalities, $\zeta$.}
    \label{fig:f3}
\end{figure}

\begin{figure}[ht!]
 \centering
  \includegraphics[width=0.48\textwidth,angle=0]{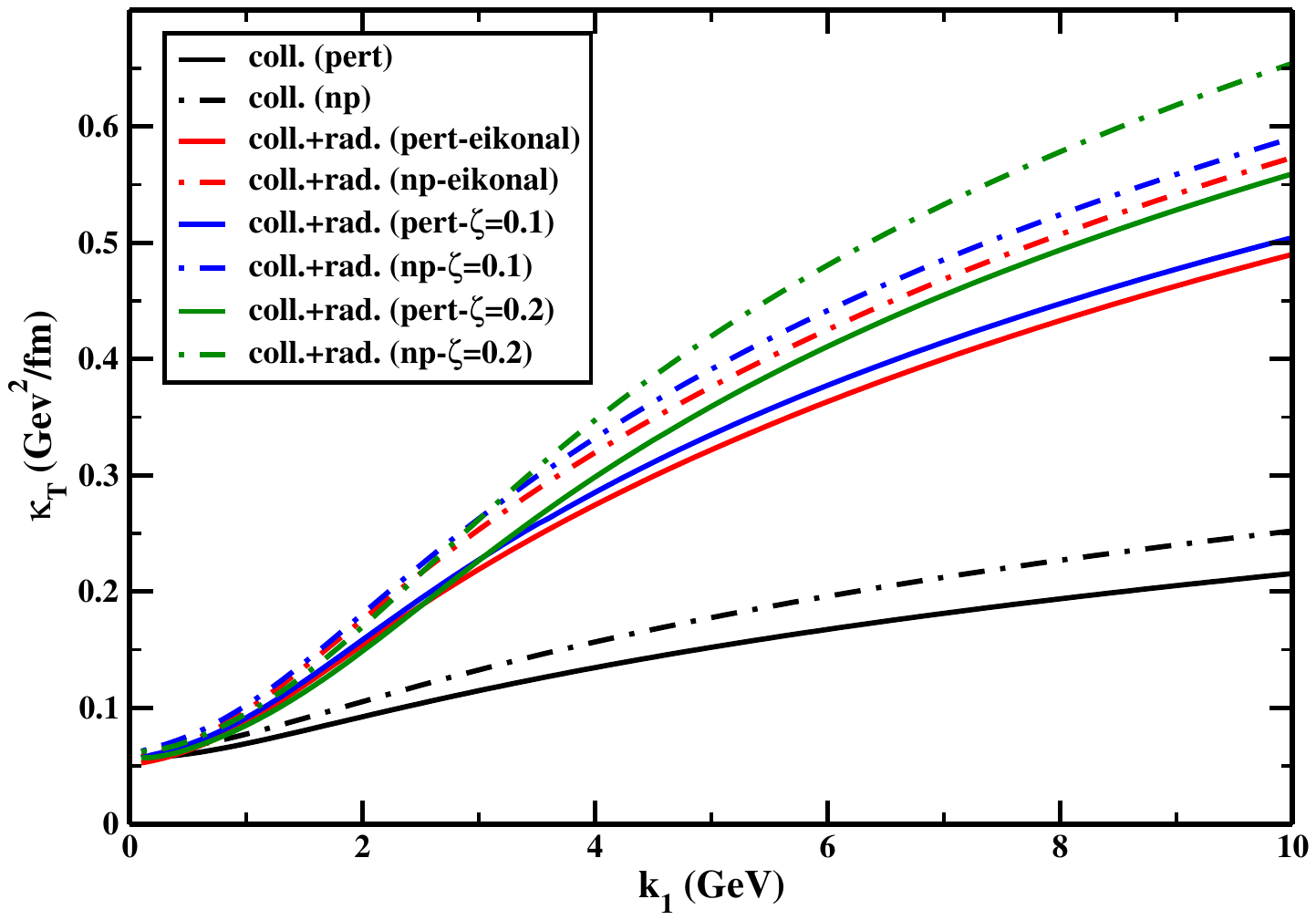}
   \caption{Variation of the transverse diffusion coefficient, $\kappa_T$, of charm quark propagating in a medium having 500 MeV temperature with charm momentum for various eikonalities, $\zeta$.}
    \label{fig:f4}
\end{figure}

Figures.~\ref{fig:f3} and \ref{fig:f4} show the variations of the transverse diffusion coefficient, $\kappa_T$, of the charm quark with temperature and charm momentum, respectively. Like the drag coefficient, one can conclude similar inferences in this case regarding the effect of the non-perturbative regime in the radiative process and the non-eikonality of the charm quark on $\kappa_T$. One can easily verify that we have reproduced the results obtained in Ref.~\cite{sumitprd2024} for the HQ transport coefficients once the non-eikonality parameter, $\zeta=0$.

\begin{figure}[ht!]
 \centering
  \includegraphics[width=0.48\textwidth,angle=0]{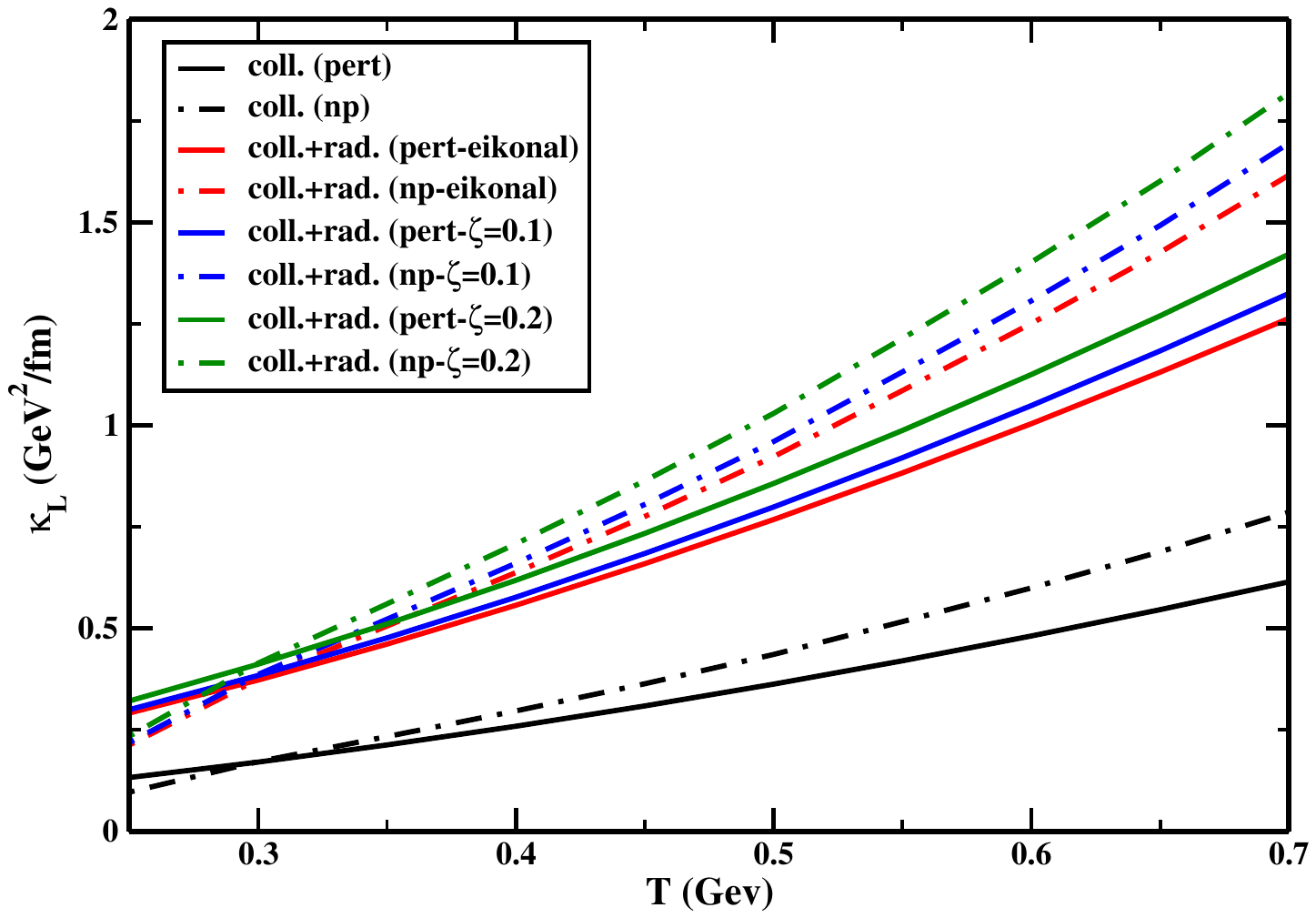}
   \caption{Variation of the longitudinal diffusion coefficient, $\kappa_L$, of charm having momentum 5 GeV with temperature of the medium for different eikonalities, $\zeta$.}
    \label{fig:f5}
\end{figure}

\begin{figure}[ht!]
 \centering
  \includegraphics[width=0.48\textwidth,angle=0]{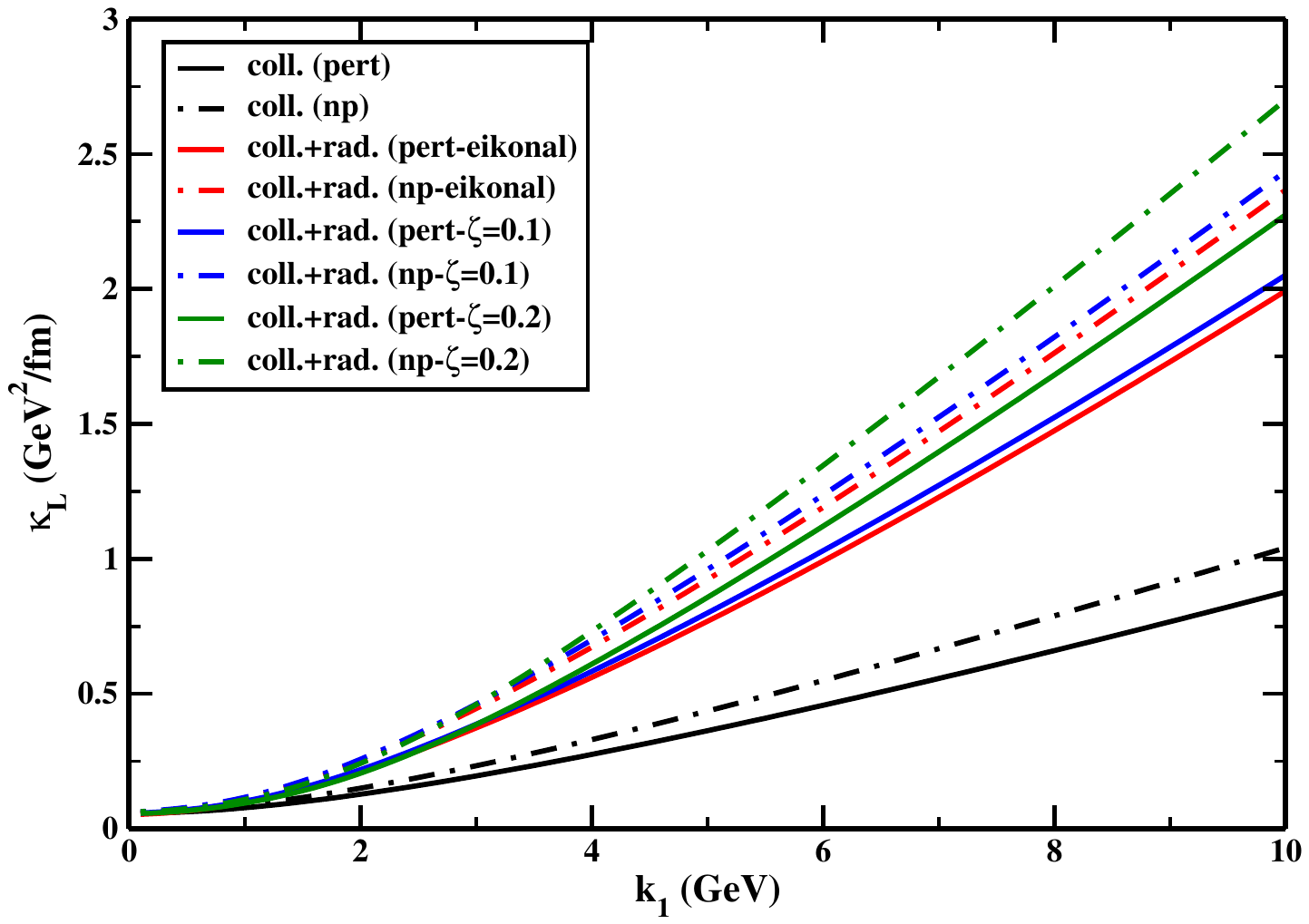}
   \caption{Variation of the longitudinal diffusion coefficient, $\kappa_L$, of charm quark propagating in a medium having 500 MeV temperature with charm momentum for various eikonalities, $\zeta$.}
    \label{fig:f6}
\end{figure}

The importance of the incorporation of the non-perturbative effect and the non-eikonality of the charm quark in the calculation of the radiative processes is explicitly manifest in the plots of the longitudinal diffusion coefficient shown by the Figs.~\ref{fig:f5} and \ref{fig:f6}. Whereas the drag coefficient signifies the depletion of the peak value in the charm quark momentum distribution, the diffusion coefficients connote the flattening of the width of the momentum distribution in the transverse and longitudinal directions to the direction of the propagation of the charm quark. If the interaction of the charm quark with the medium particles is non-eikonal, the broadening of the width of the distribution is more probable which contributes to an amplified diffusion coefficient.

\section{\bf Summary and Conclusion}
    
The present work has been dedicated to calculate the drag and diffusion coefficients of the HQ interacting with the light quarks/anti-quarks and gluons of the thermal QCD medium via both the collision and the gluon radiation using the non-perturbative Gribov propagator for mediating gluons. We emphasize on relaxing the eikonal approximation of the HQ radiating gluons in the first order of opacity and try to observe the effect of the degree of non-eikonality on the transport coefficients, both perturbatively and non-perturbatively. Every time the HQ falls off trail after a collision by a measurable quantity, $\zeta$, the spectrum of the radiated gluon contribute to an enhanced effective (collisional+radiative) transport coefficient. We see this rise of the transport coefficient for both the perturbative and the non-perturbative cases. On account of the strongly interacting medium, the non-perturbative effects increases the magnitudes of the transport coefficients even more. 

The transport of the HQ is governed by the equation of motion of a generic Brownian particle, \textit{i.e.} Langevin (L) or Fokker-Planck  (FP) equation. The transport coefficients of the HQ play pivotal role as the inputs to the L or FP equation that are solved to determine the theoretical estimation of the observables such as $R_{AA}$ or $v_2$ of the HQ. Consequently, the kinematics and dynamics of the physical interactions going into the evaluation of the transport coefficients propagates to the experimental observables, directly. We perform explicit calculations of the non-perturbative drag and transverse and longitudinal diffusion coefficients of the HQ undergoing collision as well as non-eikonal gluon radiation and contrast the results with those for the perturbative and eikonal gluon radiation cases. 

We surmise that the inclusion of the non-eikonality of the HQ emitting gluons after it collides with the medium partons enhances the effective drag and diffusion coefficients of the HQ. We expect to shed more light on the behaviour of the experimental quantities such as $R_{AA}$ and $v_2$ with respect to the momentum of the HQ with the help of this new exploration. 

We hope to continue our quest on the non-perturbative gluon radiation with more rigor by advancing beyond the single gluon emission and other approximations mentioned in the Sec.\ref{approx}, leading to a better understanding of the relevant experimental data.

\section*{Acknowledgements}

SM and NS acknowledge the support from Indian Institute of Science Education and Research, Berhampur, Odissa, India. LK acknowledges the support from NISER, Bhubaneswar, India and financial support from the research Project No. SR/MF/PS-02/2021-PU (E-37120) of the Department of Science and Technology, Government of India. 

\bibliography{npHQrad.bib}

\end{document}